# Comparative survey: People detection, tracking and multi-sensor Fusion in video sequence

Hiliwi Leake KIDANE

Laboratory Le2i, Universite Bourgogne - Franche-Comte, 21000 Dijon, France,

hiliwi-leake.kidane@u-bourgogne.fr

Abstract

Tracking people in a video sequence is one of the fields of interest in computer vision. It has broad applications in motion capture and surveillance. However, due to the complexity of human dynamic structure, detecting and tracking are not straightforward. Consequently, different detection and tracking techniques with different applications and performance have been developed. To minimize the noise between the prediction and measurement during tracking, Kalman filter has been used as a filtering technique. At the same time, in most cases, detection and tracking results from a single sensor is not enough to detect and track a person. To avoid this problem, using a multi-sensor fusion technique is indispensable. In this paper a comparative survey of detection, tracking and multi-sensor fusion methods are presented.

Keywords
Detection, Tracking, Kalman filter, multi-sensor fusion.

## 1 INTRODUCTION

Tracking people as they move through video sequences is one of the most basic and most important tasks in computer vision [1], [2], [3]. It has numerous applications in surveillance vehicle tracking, motion capture, human-computer interaction, robotics etc. Due to the complexity of human dynamic structures, the detection is not easy and straightforward. As a result, there are deferent detection and tracking techniques with different applications and capacities.

The detection results may not represent actual values as always there are artifacts introduced during video capturing. Consequently, it is important to a filter in order to estimate the measured signals. Kalman filter is used in this paper as it is an optimal signal estimator that provides the estimation of a signal in a noise [4], [5]. To minimize the covariance of the above two techniques, a Kalman filter is used as filtering method. Finally, two methods of multi-sensor data fusion are implemented.

For parallel and improved computational gain, some researchers in [6], [7], [8], [9] proposed hardware implementations of HoG and Camshaft. Similarly, hardware-based data fusion algorithms are implemented in [4]. However, due to the complexity of human dynamic structure, there are numerous algorithm developed for detection, tracking, and multi-sensor data fusion. These different techniques provide different performances depending on the application. In this paper, comparative survey of HOG and CAMShift detectors, Kalman filter based tracking and multi-sensor fusion methods are presented.

## 2 DETECTION

The People detection is the first step in different tracking applications [10]. It is the process of determining the presence of a person in video frames and determines the location and size of the person. There are different human detection algorithms like Histograms of Oriented Gradient (HOG)[11] and CAMshift[12], Viola and Jones [13]. However, as HOG and CAMshift has proposed by many authors, in this paper comparison of HOG and CAMshift will be presented.

### 2.1 HOG for People Detection

HOG object or people detection technique works based on locally contrast normalized histogram of orientation gradients. Local object appearance and shape are characterized by the distribution of local intensity gradients or edge directions [11]. The HOG detection method is proposed used for people detecting people in video sequence in [3], [11], [14], [15] and [16]. HOG works based on local contrast normalized histogram of orientation gradients. Local object appearance and shape are characterized by the distribution of local intensity gradients or edge directions [11].

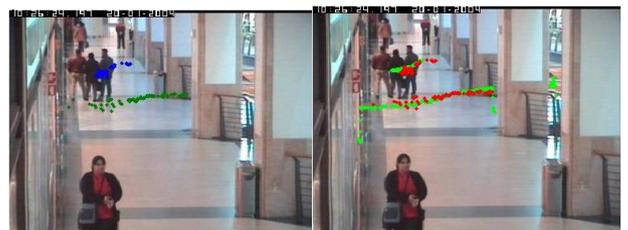

(a) HOG detection      (b) HOG(red) Vs GT(yellow)

**Figure 1** HOG detection and corresponding GT.

## 2.2 CAMshift for People Detection

The CAMshift algorithm has been used as detection method for people tracking [17], [18], [19], [20] and [21]. The algorithm applies color data to probability distribution, especially hue data in HSV. The color probability distribution map called back projected image is made and used for tracking operations [12]. As this algorithm finds the color probability distribution to track the object, applying directly on the given video did not give the expected result. This is due to the fact that the color distribution may be dominated by the static background. To minimize the effect of the static background, a background subtraction can be used before applying the CAMshift. The result of CAMshift detection is given in Fig 2. Out of the six people tracked in the ground truth, only one person is detected. This is because of the fact that when applying background subtraction, people in very slow motion will be treated as static object and removed. In this specific lab threshold is used to remove the noise coming from the people moving slowly. From the actual video, the person going out of the shop is moving fast comparing with the other people in the video. The question is why the centroids of the CAMshift detection are lower than the centroids of the ground truth. This may be due to the high variation in the half lower part of the person's body including his hands and feet.

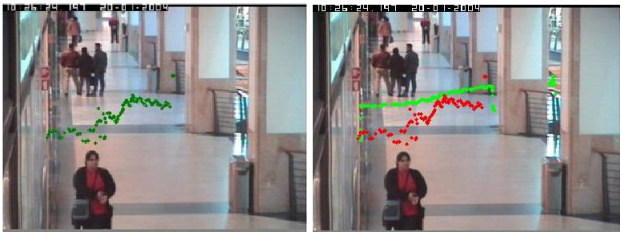

(a) CAMshift detection    (b) CAMshift(red) Vs GT(yellow)

**Figure 2** CAMshift detection and corresponding GT.

## 2.3 Pros and cons

The HOG detectors gives better results both in terms of the number of detected peoples and accuracy of the tracked two people with respect the ground truth. The limitation of HOG is computation speed. As the algorithm works based on local normalized gradient, it is very slow. On the other hand, as the CAMshift was applied on background subtracted image, it tracks only the person in high motion as there is high variation in the pixel intensity (hue) around this region. The CAMshift is very fast but not gives accurate tracking result. Both the above detectors are not invariant to occlusion. The MSE of both detectors is computed with respect to the ground truth tracking for the man going out from the shop as shown in Fig. 3 and Table III. The result shows that the HOG detection method provides better performance than the CAMshift.

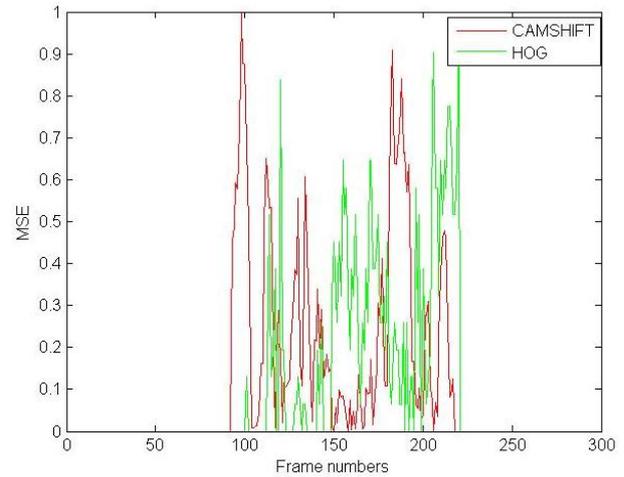

**Figure 3** MSE of HOG and CAMshiftT with respect to GT.

| HOG | CAMshift |
|---|---|
| 17.03 | 20.07 |

**Table 1** Quantitative MSE of the HOG and CAMshift

Even though HOG detectors give better results both in terms of the number of detected people and accuracy of the position of the detected people, it has low speed and is not robust for occlusion. To overcome the problem related to computational speed, Authors in [22], [6], [8], [15] have proposed hardware based implementation of HOG for people tracking. However, there are always some outliers and variance in the detection and can be improved by applying a filter. Kalman filter [23] has been proposed as an efficient filtering technique to predict the next detection and improve the overall tracking. One way to minimize the problems related to occlusion is to use multi-camera and use the multi-camera fusion techniques as proposed in [16].

## 3  TRACKING

Tracking is the process of finding trajectories of objects over time from a given sequence of images. There are two approaches of tracking methods. The first is deterministic methods where tracking is performed by an iterative search for a local maximum of similarity measure between a template of the target and the current frame. The second tracking method is a probabilistic method which uses a state-space representation of a moving object to model its underlying dynamics. In this method, the tracking problem is viewed as a Bayesian interference problem. Examples are Kalman filter and particle filters.

The advantage of the second tracking method is, it avoids the bulk work for segmentation and searching in the full image. There are three key steps in this process, prediction, matching (data association) and correction (updating). Tracking involves two basic steps motion and matching. Motion predicts the limited search region in which the tracked object most likely to be in the next frame. Matching identifies the object within the designed search region.

Combining the detection methods with estimation filters can easily improve the tracking result. The Kalman filter has been used as estimation method for object and people tracking [24], [25], [26]. Authors in [23] have proposed a combination of HOG and Kalman filter for efficient human detection and tracking in a video.

The detections can be improved by applying a Kalman filtered modeled using the standard equations of motions for constant velocity and acceleration. Considering the centroid of the tracking as point object in motion from one frame to the next frames, the displacement can be expressed using equations 1 and 2 for constant velocity and acceleration in 1D.

$$r = r_0 + vt \tag{1}$$

$$r = r_0 + vt + (at^2)/2 \tag{2}$$

Since the tracking is in 2D, the above equation can be easily adopted to fit with the 2D motion model by decomposing the centroid r into its x and y component. In this case of constant velocity, we will have 4x4 transition matrix representing the displacement and velocity in x and y. At the same time for constant acceleration, the transition matrix will be 6x6 representing displacement, velocity and acceleration in x and y directions. The state transition and observation matrices for constant velocity model are given in equation 3 and 4.

$$X_{(cu)} = \begin{pmatrix} 1 & 0 & 1 & 0 \\ 0 & 1 & 0 & 1 \\ 0 & 0 & 1 & 0 \\ 0 & 0 & 0 & 1 \end{pmatrix} \tag{3}$$

$$Z_{(cu)} = \begin{pmatrix} 1 & 0 & 0 & 0 \\ 0 & 1 & 0 & 0 \end{pmatrix} \tag{4}$$

A 6x6 matrix for state transition and 2x6 matrix for observation are used in the acceleration model in a similar manner as the equations of the constant velocity model. Process noise and the measurement noise matrices are assumed to be identity.

The second order Kalman filter (position, velocity, and acceleration) is used to model the motion of each person in the scene. The Kalman filter works in two stages: prediction and correction. Taking the HOG detection and CAMshift as a measurement/observation for the Kalman filter, the next possible detection is predicted using the state transition matrix derived from the equation of motion model. The challenge comes when the person under tracking is missed after some time. In this case there will be only prediction without any update. The difference between constant velocity and acceleration comes in this step. From our experiment, the acceleration model gives better prediction result when the person is detected. On the other hand, the constant velocity model performs better in prediction when the person is missed in the middle.

The overall MSE of both detectors before and after Kalman filter is given in Table 2. The result shows that Kalman filter improves the tracking.

| Technique | Before filtering | After filtering |
|---|---|---|
| HOG | 17.03 | 14.68 |
| CAMshift | 20.07 | 17.21 |

**Table 2** MSE Before and after Kalman filtering

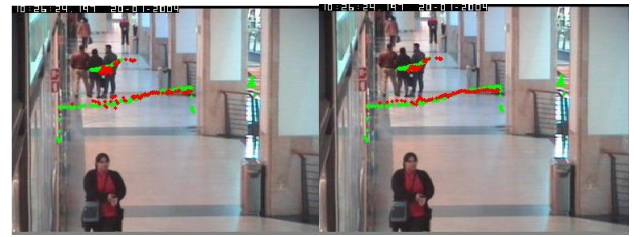

(a) (b)

**Figure 4** HOG detection a) Before Kalman filter (yellow=GT, red=HOG) b) After Kalman filter (yellow=GT, red=HOG).

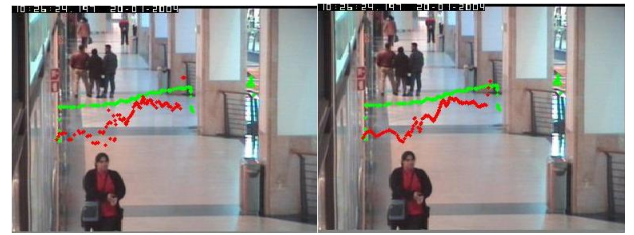

(a) Before filter(green=GT, red=CAM) (b) After filter (green=GT, red=CAM)

**Figure 5** CAMshift detection before and after Kalman filter.

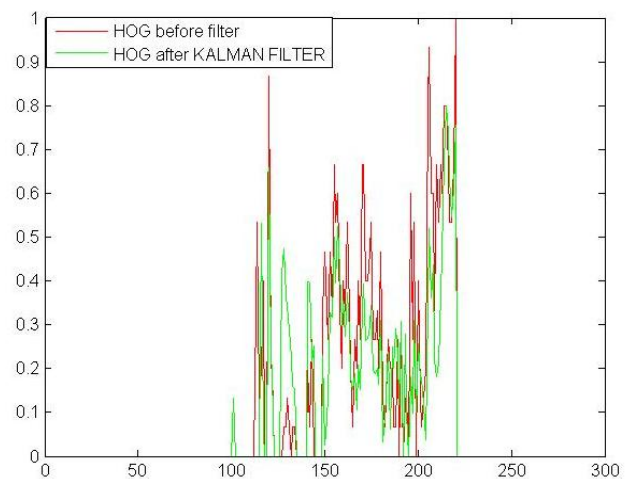

**Figure 6** MSE of HOG before and after Kalman filter.

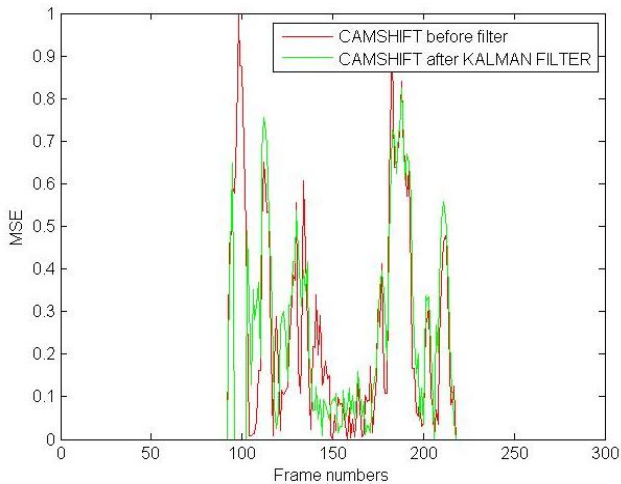

**Figure 7** MSE of CAMshift before and after Kalman filter.

## 4 DATA FUSION

Data fusion [27] [28] is the process of combining information from a number of different sources to improve the certainty and accuracy of the information. Tracking of moving people in a video can be improved by fusing tracking of different camera capturing the same person from different angles[29], [30] as shown Fig. 8. Two techniques of multi-sensor data fusion called weighted sum and Winner-take-all are compared in this paper. The first fusing method works by taking a weighted sum of the filtered points. The second fusion method proposed works based on the rule winner-take-all. It considers the filtered point with high value as the correct answer.

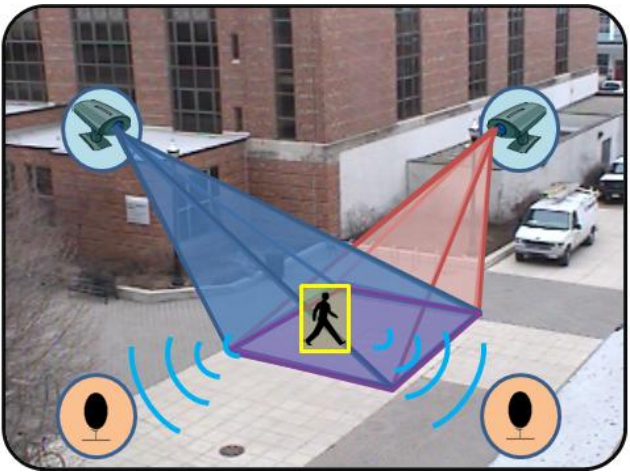

**Figure 8** multi-sensors fusion.

### 4.1 Weighted sum

The weighted sum is a linear combination method [31] of multi-sensor data fusion. The weights of each sensor depend on their tracking contribution of the target person. The performance of this fusion method is highly dependent on the weight selection. Normally when the contribution of the cameras involved in the detection of the person or the number of frames where the person is detected in all cameras are equal, a simple averaging method can be used to get the fused tracking result.

To fuse the tracking of the person from two camera using the weighted sum method on the base plane, first the Kalman filter is applied to minimize the error compared with the ground truth before projecting the tracking results. Though the result is not perfect as the ground truth, it is assumed that the deviation between GT and HOG is reduced after Kalman filter. After projecting the filtered tracking points from the two cameras, a weighted sum is used. The key point here is, the weights given to each camera tracking result. Normally when the contribution of the two cameras to the detection of the person or the number of frames where the person is detected in both cameras is equal, we can use a simple averaging method which gives equal weight to both cameras tracking result. But in this specific video used, the corridor camera has more contribution than the front. As a result, a biased weight is used to give some advance to the corridor camera. In this experiment, 0.8 by 0.2 weights are used for the corridor and front camera respectively. Since the person disappears from the front camera after some times while staying in the corridor camera, a threshold is used to give a full weight to the corridor camera when the person is missed for consecutive frames in the front camera. Otherwise the overall result will be affected due to the prediction without update in the front camera. As the main target of this practical work is to compare with the ground truth, both the ground truth and the tracking are projected in to the base plane. But unlike the HOG detection, to fuse the GT, a simple averaging method is used when the person detected from both sides and if it is detected only in one side the result of the single camera is considered without averaging. The fusion result of this method showed that the MSE is reduced to 13.23 from 17.03..

### 4.2 Winner-take-all

Winner-take-all [32] is a fusion method where the entire sensor which gives better result is selected. This multi-sensor fusion method is important when the discrepancy contributions of the sensors are different. It helps to avoid the contribution of an unhealthy sensor to the overall tracking result. In addition, when the person in a track is disappeared from the sensor after some time, there will be only prediction without update/detection. Consequently, the tracking points from this sensor are unhealthy and it has to be ignored and the Winner take-all method helps to select the one with the highest score.

### 4.3 Pros and cons of Fusion Methods

The weighed sum fusion method is easy and straight forward to implement and gives good result as shown in Fig. 9. If the detection from both cameras is almost similar and comparable, it is possible to achieve better result by simply taking average of the two points. But if the detection from both camera is not equal like the front camera which is dominated by wrong prediction without measurement/update while the corridor camera gives right prediction and measurement, then the performance of this method will be highly dependent on the selection of appropriate weight. Although it is possible to improve the performance by tuning to the weight which gives better, the selection of weight is not robust. This is the main drawback of this method. The other point to consider is even after accurate choice of appropriate weight, there will be some outliers in the result. This is due to the fact that if detection is missed in one of the camera, the method is taking a single detection times its weight and as the weight is less than 1, the result will be affected to somehow. This problem is also reduced by using some threshold to switch into use of the detection as it is in seated of multiplying by weight when a detection is missed for consecutive 3 in one of the camera.

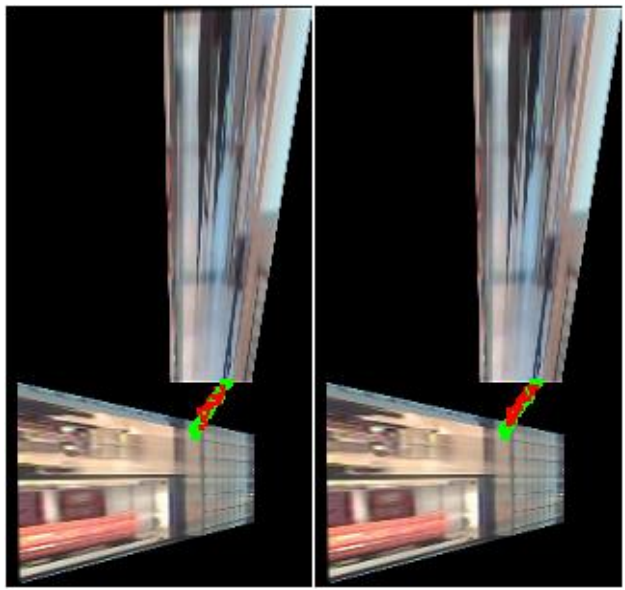

**Figure 9** Fusion results for both methods (green=GT, red=HOG).

The Winner take all method gives better fusion result by biasing into the detection which gives better score. The important part of this method is when detection is missed from one cameras, it will take all the information from the other camera. This gives better result as the effect of the existing detection is not reduced by multiplying with some fraction weight like the weighted sum fusion method. The drawback of this method is neglecting the contribution of the second detection when both cameras detect a person. Though the choice of the point is based on score, neglecting the effect of the second detection at all is not good as there may have some discrepancy the selected one.

The overall MSE starting from detection to fusion is shown in table 3. From the result it can be concluded that the fusion improves the overall tracking.

| HOG | HOG+Kalman | Weighted sum | Winner take all |
|---|---|---|---|
| 17.03 | 14.68 | 13.23 | 12.91 |

**Table 2** Camparision of MSE of HOG, Kalman and Fusion methods

## 5 CONCLUSION

In this paper, comparisons of detection, tracking and multisensory fusion methods for people tracking are presented. Two detection methods, HOG and CAMshift are compared with the ground truth and measured the MSE. Based on the MSE result, the HOG detector provides better result. Then, a Kalman filter is applied on the tracking to improve the tracking results by the detectors HOG and CAMshift. The Kalman filter gives better tracking result reducing the MSE by a quarter.

Finally, two fusion methods are implemented and compared. The filtered detection result of the same person from two cameras is fused by projecting into a base plane. The first method is based on weighted sum and the second is based on winner take all rule. Both fusion methods give batter result in reducing the MSE. The contributions of the two cameras used in this experiment were not balanced. As a result, a biased weights are used to give more priority to the one contributes more. But if the position of the cameras is adjust to a position where both can contribute equally, the MSE can be reduced more to achieve maximum tracking results.

.

## 6 REFERENCES

[1] F. da Silva Guizi and C. S. Kurashima, "Real-time people detection and tracking using 3d depth estimation," in 2016 IEEE International Symposium on Consumer Electronics (ISCE), Sept 2016, pp. 39–40.

[2] D. L. Siqueira and A. M. C. Machado, "People detection and tracking in low frame-rate dynamic scenes," IEEE Latin America Transactions, vol. 14, no. 4, pp. 1966–1971, April 2016.

[3] Y. Zheng, A. Mahabalagiri, and S. Velipasalar, "Detection of moving people with mobile cameras by fast motion segmentation," in 2013 Seventh International Conference on Distributed Smart Cameras (ICDSC), Oct 2013, pp. 1–6.

[4] A. H. G. Al-Dhaher, E. A. Farsi, and D. Mackesy, "Data fusion architecture - an fpga implementation," in 2005 IEEE Instrumentation and Measurement Technology Conference Proceedings, vol. 3, May 2005, pp. 1985–1990.

[5] M. S. Grewal and A. P. Andrews, Kalman Filtering: Theory and Practice with MATLAB, 4th ed. Wiley-IEEE Press, 2014.

[6] P. Y. Chen, C. C. Huang, C. Y. Lien, and Y. H. Tsai, "An efficient hardware implementation of hog feature extraction for human detection," IEEE Transactions on Intelligent Transportation Systems, vol. 15, no. 2, pp. 656–662, April 2014.

[7] S. Lee, H. Son, J. C. Choi, and K. Min, "Hog feature extractor circuit for real-time human and vehicle

detection," in TENCON 2012 IEEE Region 10 Conference, Nov 2012, pp. 1–5.

[8] M. Hemmati, M. Biglari-Abhari, S. Berber, and S. Niar, "Hog feature extractor hardware accelerator for real-time pedestrian detection," in 2014 17th Euromicro Conference on Digital System Design, Aug 2014, pp. 543–550.

[9] S. Wong and J. Collins, "A proposed fpga architecture for real-time object tracking using commodity sensors," in 2012 19th International Conference on Mechatronics and Machine Vision in Practice (M2VIP), Nov 2012, pp. 156–161.

[10] T. Santhanam, C. P. Sumathi, and S. Gomathi, "A survey of techniques for human detection in static images," in Proceedings of the Second International Conference on Computational Science, Engineering and Information Technology, ser. CCSEIT '12. New York, NY, USA: ACM, 2012, pp. 328–336. [Online]. Available: http://doi.acm.org/10.1145/2393216.2393272

[11] N. Dalal and B. Triggs, "Histograms of oriented gradients for human detection," in 2005 IEEE Computer Society Conference on Computer Vision and Pattern Recognition (CVPR'05), vol. 1, June 2005, pp. 886–893 vol. 1.

[12] R. Araki, S. Gohshi, and T. Ikenaga, "Real-time both hands tracking using camshift with motion mask and probability reduction by motion prediction," in Proceedings of The 2012 Asia Pacific Signal and Information Processing Association Annual Summit and Conference, Dec 2012, pp. 1–4.

[13] P. Viola and M. Jones, "Rapid object detection using a boosted cascade of simple features," in Proceedings of the 2001 IEEE Computer Society Conference on Computer Vision and Pattern Recognition. CVPR 2001, vol. 1, 2001, pp. I–511–I–518 vol.1.

[14] S. A. Chowdhury, M. N. Uddin, M. M. S. Kowsar, and K. Deb, "Occlusion handling and human detection based on histogram of oriented gradients for automatic video surveillance," in 2016 International Conference on Innovations in Science, Engineering and Technology (ICISET), Oct 2016, pp. 1–4.

[15] S. F. Hsiao, J. M. Chan, and C. H. Wang, "Hardware design of histograms of oriented gradients based on local binary pattern and binarization," in 2016 IEEE Asia Pacific Conference on Circuits and Systems (APCCAS), Oct 2016, pp. 433–435.

[16] M. Wang, Y. Dai, Y. Liu, and T. Yanbing, "Feature-level image sequence fusion based on histograms of oriented gradients," in 2010 3rd International Conference on Computer Science and Information Technology, vol. 9, July 2010, pp. 265–269.

[17] S. Boubou, A. Kouno, and E. Suzuki, "Implementing camshift on a mobile robot for person tracking and pursuit," in 2011 IEEE 11[th] International Conference on Data Mining Workshops, Dec 2011, pp. 682–688.

[18] S. Huang and J. Hong, "Moving object tracking system based on camshift and kalman filter," in 2011 International Conference on Consumer Electronics, Communications and Networks (CECNet), April 2011, pp. 1423–1426.

[19] A. Ganoun, N. Ould-Dris, and R. Canals, "Tracking system using camshift and feature points," in 2006 14th European Signal Processing Conference, Sept 2006, pp. 1–5.

[20] L. Wang, C. Xiu, and S. Wei, "Improved camshift algorithm based on weighted histogram model," in The 27th Chinese Control and Decision Conference (2015 CCDC), May 2015, pp. 5795–5798.

[21] C. Xiu and F. Ba, "Target tracking based on the improved camshaft method," in 2016 Chinese Control and Decision Conference (CCDC), May 2016, pp. 3600–3604.

[22] R. Kadota, H. Sugano, M. Hiromoto, H. Ochi, R. Miyamoto, and Y. Nakamura, "Hardware architecture for hog feature extraction," in 2009 Fifth International Conference on Intelligent Information Hiding and Multimedia Signal Processing, Sept 2009, pp. 1330–1333.

[23] C. Li, L. Guo, and Y. Hu, "A new method combining hog and kalman filter for video-based human detection and tracking," in 2010 3[rd] International Congress on Image and Signal Processing, vol. 1, Oct 2010, pp. 290–293.

[24] C. Suliman, C. Cruceru, and F. Moldoveanu, "Kalman filter based tracking in an video surveillance system," in Advances in Electrical and Computer Engineering, vol. 10, 2010, pp. 30–34.

[25] B. Zheng, X. Xu, Y. Dai, and Y. Lu, "Object tracking algorithm based on combination of dynamic template matching and kalman filter," in 2012 4th International Conference on Intelligent Human-Machine Systems and Cybernetics, vol. 2, Aug 2012, pp. 136–139.

[26] A. Kulkarni, M. Vargantwar, and S. Virulkar, "Video based tracking and optimization using mean-shift, kalman filter and swarm intelligence," in 2015 International Conference on Computing Communication Control and Automation, Feb 2015, pp. 629–633.

[27] A. Kraussling and D. Schulz, "Data fusion for person identification in people tracking," in 2008 11th International Conference on Information Fusion, June 2008, pp. 1–8.

[28] B. Khaleghi, A. Khamis, F. O. Karray, and S. N. Razavi, "Multisensor data fusion: A review of the state-of-the-art," Information Fusion, vol. 14, no. 1, pp. 28 – 44, 2013. [Online]. Available: http://www.sciencedirect.com/science/article/pii/S1566253511000558

[29] N. T. Siebel and S. J. Maybank, "Fusion of multiple tracking algorithms for robust people tracking," in

Proceedings of the 7[th] European Conference on Computer Vision-Part IV, ser. ECCV '02. London, UK, UK: Springer-Verlag, 2002, pp. 373–387. [Online]. Available: http://dl.acm.org/citation.cfm?id=645318.649351

[30] K. Toyama and G. D. Hager, "Tracker fusion for robustness in visual feature tracking," pp. 38–49, 1995. [Online]. Available: http://dx.doi.org/10.1117/12.220965

[31] Y. Zhou and H. Leung, "Minimum entropy approach for multisensory data fusion," in Proceedings of the IEEE Signal Processing Workshop on Higher-Order Statistics, Jul 1997, pp. 336–339.

[32] H. Osman and S. D. Blostein, "Probabilistic winner-take-all segmentation of images with application to ship detection," IEEE Transactions on Systems, Man, and Cybernetics, Part B (Cybernetics), vol. 30, no. 3, pp. 485–490, Jun 2000.